\documentclass[aps,prstab,reprint,groupedaddress,nofootinbib]{revtex4-1}

\usepackage{graphicx}% Include figure files
\usepackage{dcolumn}% Align table columns on decimal point
\usepackage{bm}% bold math
\usepackage{amsmath}
\usepackage{units}
\usepackage{lipsum}  

\usepackage[colorinlistoftodos]{todonotes}
\usepackage[caption=false]{subfig}

\usepackage{hyphenat}
\usepackage{color}

\begin{document}

\global\long\def\d{\mathrm{d}}
\global\long\def\en{\varepsilon_{0}}
\global\long\def\e{\mathrm{e}}
\global\long\def\ii{\mathbf{\imath}}

%\preprint{}


\title{Systematic Studies of the Micro-Bunching Instability at Very Low Bunch Charges}



\author{Miriam Brosi}\email{miriam.brosi@kit.edu}
\author{Johannes L. Steinmann}
\author{Edmund Blomley}
\author{Erik Br\"undermann}
%\author{Nicole Hiller}\altaffiliation{now at PSI, Villigen, Switzerland}
\author{Julian Gethmann}
\author{Benjamin Kehrer}
\author{Yves-Laurent Mathis}
%\author{Michael J.\ Nasse}
\author{Alexander Papash}
\author{Manuel Schedler}\altaffiliation{now at Varian PT, Troisdorf, Germany}
\author{Patrik Sch\"onfeldt}
\author{Marcel Schuh}
\author{Markus Schwarz}\altaffiliation{now at CERN, Geneva, Switzerland}
\author{Anke-Susanne M\"uller}
\affiliation{LAS and IBPT, Karlsruhe Institute of Technology, Germany}
\author{Michele Caselle}
\author{Lorenzo Rota}
\author{Marc Weber}
\affiliation{IPE, Karlsruhe Institute of Technology, Germany}
\author{Peter Kuske}
\affiliation{Humboldt-Innovation GmbH and Helmholtz-Zentrum Berlin, Germany}


\date{\today}

\begin{abstract}
 At KARA, the KArlsruhe Research Accelerator of the KIT synchrotron,
 %  the accelerator test facility and synchrotron light source ANKA at the Karlsruhe Institute of Technology (KIT)
 the so called short bunch operation mode allows the reduction of the bunch length down to a few picoseconds.  
The micro-bunching instability resulting from the high degree of longitudinal compression leads to fluctuations in the emitted THz radiation, referred to as bursting. 
For extremely compressed bunches at KARA, bursting occurs not only in one but in two different bunch-current ranges that
are separated by a stable region. 
This work presents measurements of the bursting behavior in both regimes. Good agreement is found between data and numerical solutions of 
the Vlasov-Fokker-Planck equation.
\end{abstract}

\pacs{no PACS numbers yet}
%\keywords{}

\maketitle
%\tableofcontents
\section{Introduction}
The KArlsruhe Research Accelerator (KARA) is the storage ring of the accelerator test facility and synchrotron light source of the Karlsruhe Institute of Technology (KIT) in Germany.
%ANKA is a synchrotron radiation source located in Karlsruhe, Germany, and is operated by the Karlsruhe Institute of Technology. 
%The storage ring with a circumference of  \unit[110.4]{m} can be operated at energies ranging from  \unit[0.5]{GeV} to  \unit[2.5]{GeV}. 
A special short-bunch operation mode at  \unit[1.3]{GeV} allows the reduction of the momentum compaction factor and therefore reduces the electron bunch length down to a few picoseconds. 
The bunch-by-bunch feedback system~\cite{hertle_ipac14} enables custom filling patterns from a single bunch to complex multi bunch patterns.

Coherent synchrotron radiation (CSR) is emitted for wavelengths in the order of or longer than the emitting structure. In the short-bunch operation mode the effect of CSR plays an important role in the beam dynamics. 
The compressed bunch length of a few picoseconds leads to the emission of CSR in the low THz frequency range, that is related to a modulation of the longitudinal phase space due to the CSR impedance. 
This modulation leads to substructures in the longitudinal particle distribution and is referred to as micro-bunching~\cite{stupakov2002}. 
The time varying substructures lead to strong fluctuations of the emitted power in the THz range, the so-called bursting. 
The minimum bunch-current at which this phenomenon occurs is called the bursting threshold. It depends strongly on the natural bunch length and therefore on various machine parameters~\cite{Bane_cai_stupakov2010}. 

At KARA at KIT~\cite{asm_ipac13,brosi_ipac16} as well as at MLS~\cite{ries_ipac12}, for very short bunches, bursting was additionally observed in a region below the main bursting threshold. Indications also can be seen in measurements at Diamond (see~\cite{diamond_bursting_2012} Fig.~6) though it was not discussed further.
This instability is referred to as short bunch-length bursting (SBB) in the following.

\section{Theoretical description}\label{chap:theory}
The interaction of the electrons inside an electron bunch with their CSR radiation field, which causes the micro-bunching instability, can be described by the Vlasov-Fokker-Planck (VFP) equation~\cite{stupakov2002}. 
The solution depends on how the effects of the conductive beam pipe are taken into account as boundary conditions for the emitted electromagnetic field. 
In this contribution, the model, to which the measurements will be compared to, treats the beam pipe as a pair of perfectly conducting parallel plates with a distance of $2h$.
The resulting solution for the main threshold of the instability was published in 2010 by Bane, Stupakov and Cai~\cite{Bane_cai_stupakov2010}. Using the dimensionless quantities $S_{\mathrm{CSR}}$ and $\Pi$ to parametrize CSR strength and shielding, the main threshold can be described by the simple linear scaling law~\cite{Bane_cai_stupakov2010}

\begin{align}
\left(S_{\mathrm{CSR}}\right)_{\mathrm{th}} &= 0.5 + 0.12\ \Pi \label{equ:threshold}\\
\mathrm{with~}
\Pi &=  \frac{\sigma_{\mathrm{z},0} R^{1/2}}{ h^{3/2}}\label{equ:pi}\\
\mathrm{and~}
S_{\mathrm{CSR}} &= \frac{I_{\mathrm{n}} R^{1/3} }{ \sigma_{\mathrm{z},0}^{4/3}}\label{equ:S_csr}.
\end{align}
Here $\sigma_{\mathrm{z},0}$ is the natural bunch length, $R$ the bending radius, $h$ half of the spacing between the parallel plates and $I_{\mathrm{n}}$ the normalized bunch current:
$$I_{\mathrm{n}} = \frac{r_{\mathrm{e}} N_{\mathrm{b}}}{2 \pi \nu_{\mathrm{s},0} \gamma \sigma_{\delta,0}}=\frac{I_{\mathrm{b}} \sigma_{\mathrm{z},0}}{\gamma \alpha_{\mathrm{c}} \sigma_{\delta,0}^{2} I_{\mathrm{A}}}$$
with $N_{\mathrm{b}}$ the number of electrons, $I_{\mathrm{b}}$ the bunch current, $r_{\mathrm{e}}$ the classical electron radius, $\nu_{\mathrm{s},0}$ the nominal synchrotron tune, $\sigma_{\delta,0}$ the natural energy spread, $\alpha_{\mathrm{c}}$ the momentum compaction factor, $\gamma$ the Lorentz factor and $I_{\mathrm{A}}$ the Alfv\'en current\footnote{Alfv\'en current $I_{\mathrm{A}}=4\pi\varepsilon_{0}m_{\mathrm{e}}c^{3}/e= \unit[17045]{\mathrm{A}}$}.
The natural bunch length is given by \cite{wiedemann_2013}
$$\sigma_{\mathrm{z},0}= \frac{\alpha_{\mathrm{c}}\, c\, \sigma_{\delta,0}}{2\pi f_{\mathrm{s}}} 
 \quad\mbox{with}\quad
f_{\mathrm{s}} = \sqrt{\frac{\alpha_{\mathrm{c}} \,f_{\mathrm{RF}} f_{\mathrm{rev}} \sqrt{e^{2}V_{\mathrm{RF}}^{2} - U_{0}^{2}}}{E\, 2 \pi}}$$
with the synchrotron frequency $f_\mathrm{s}$, the beam energy $E$, the RF frequency $f_{\mathrm{RF}}$, 
the revolution frequency $f_{\mathrm{rev}}$, the RF peak voltage $V_{\mathrm{RF}}$ and 
the radiated energy per particle and revolution $U_0$.

Equation~(\ref{equ:threshold}) was obtained from the results of a VFP-solver, based on an algorithm devised by Warnock and Ellison \cite{warnock_2000}. 
It is the result of a linear fit to the simulated thresholds. % $\left(S_{\mathrm{CSR}}\right)_{\mathrm{th}}$ when displayed as a function of $\Pi$. 
This linear scaling law fits best for high values of $\Pi$ ($\Pi > 3$), however at lower values the simulated thresholds are significantly higher than the fit. 
Interestingly, the measured thresholds are described more accurately by the fit (see~\cite{brosi_prab16}) than the thresholds obtained by the VFP simulations for these low values of $\Pi$. This is the case for the VFP simulations in~\cite{Bane_cai_stupakov2010} as well as for the ones presented in the following and can be attributed to the simplicity of the parallel plates model.
%\textcolor{blue}{This is visible again in the performed simulations presented in the following.  
%In~\cite{brosi_prab16} it was shown, that also for the lower values of $\Pi$ the scaling law describes the measured thresholds quite well. This already points to deficiencies/limitations (?) of the simple parallel plates model, as discussed later.}
%Nevertheless, it was shown in  \cite{brosi_prab16}, that it can be used also for lower values of $\Pi$ as it describes the measured thresholds quite well. \todo[inline,color=blue!30]{Die Übereinstimmung zwischen Experiment und dem falschen Skalierungsverhalten in diesem Parameterbereich ist also schon ein Hinweis auf Defizite des Modells}
%\todo[inline,color=blue!30]{fits best for $\Pi > 3$, below simulated thresholds significantly (Peter) higher....?} 

The fit, leading to Eq.~(\ref{equ:threshold}), ignores a dip around $\Pi\approx0.7$, where the calculated thresholds deviate from the simple linear scaling law~\cite{Bane_cai_stupakov2010}. 
A closer look~\cite{Bane_cai_stupakov2010} at the calculated energy spread at this dip %($\Pi=0.7$) 
reveals a second unstable region in the bunch current, with a threshold below the one expected from the linear scaling law. For the values of $\Pi$ accessible at KARA, it is separated by a stable region from the main instability starting at the threshold described by Eq.~(\ref{equ:threshold}). 
This new region corresponds to the short bunch-length bursting studied in the following.
%\textcolor{blue}{
The upper and lower limits of this additional region of instability are predicted to depend not only on the bunch length and thus the shielding parameter, but also on  $\beta=1 / \left(2 \pi \, f_{\mathrm{s}}\, \tau_{\mathrm{d}}\right)$, which relates the synchrotron frequency  $f_{\mathrm{s}}$ and the longitudinal damping time $\tau_{\mathrm{d}}$ \cite{Bane_cai_stupakov2010, Kuske_ipac13}. It is therefore termed a weak instability. %\todo[inline,color=blue!30]{because it is weak instability compared to main bursting which is a strong instability and independent of the damping time}%}

\section{Measurements}
\subsection{Measurement Principle}
The measurements presented in this paper were obtained with a broad-band quasi-optical Schottky barrier diode from ACST~\cite{acst_flyer} sensitive in the spectral range from several \unit[10]{GHz} up to \unit[2]{THz} with the peak sensitivity around \unit[80]{GHz}, which is operated at room temperature. 
The THz radiation was detected at the Infrared2 Beamline, which provides synchrotron radiation from the entry edge of a dipole magnet~\cite{Mathis2003}.

To detect the fluctuations in the emitted THz radiation for each bunch in a multi bunch filling pattern individually, the fast detector was combined with the ultra-fast DAQ system KAPTURE (KArlsruhe Pulse Taking and Ultrafast Readout Electronics)~\cite{caselle_ipac14}. 
The KAPTURE system samples the detector response to the THz pulse of each bunch at four points ~\cite{caselle_ibic14}. 
In principle KAPTURE can sample the signal continuously with the rate of the RF frequency of KARA ($\approx$  \unit[500]{MHz}). 
For this publication, the signal was recorded for every bunch at every 10th revolution during a period of one second, to limit the acquired data volume. 

\begin{figure}[!t]
   \includegraphics*[width=0.5\textwidth]{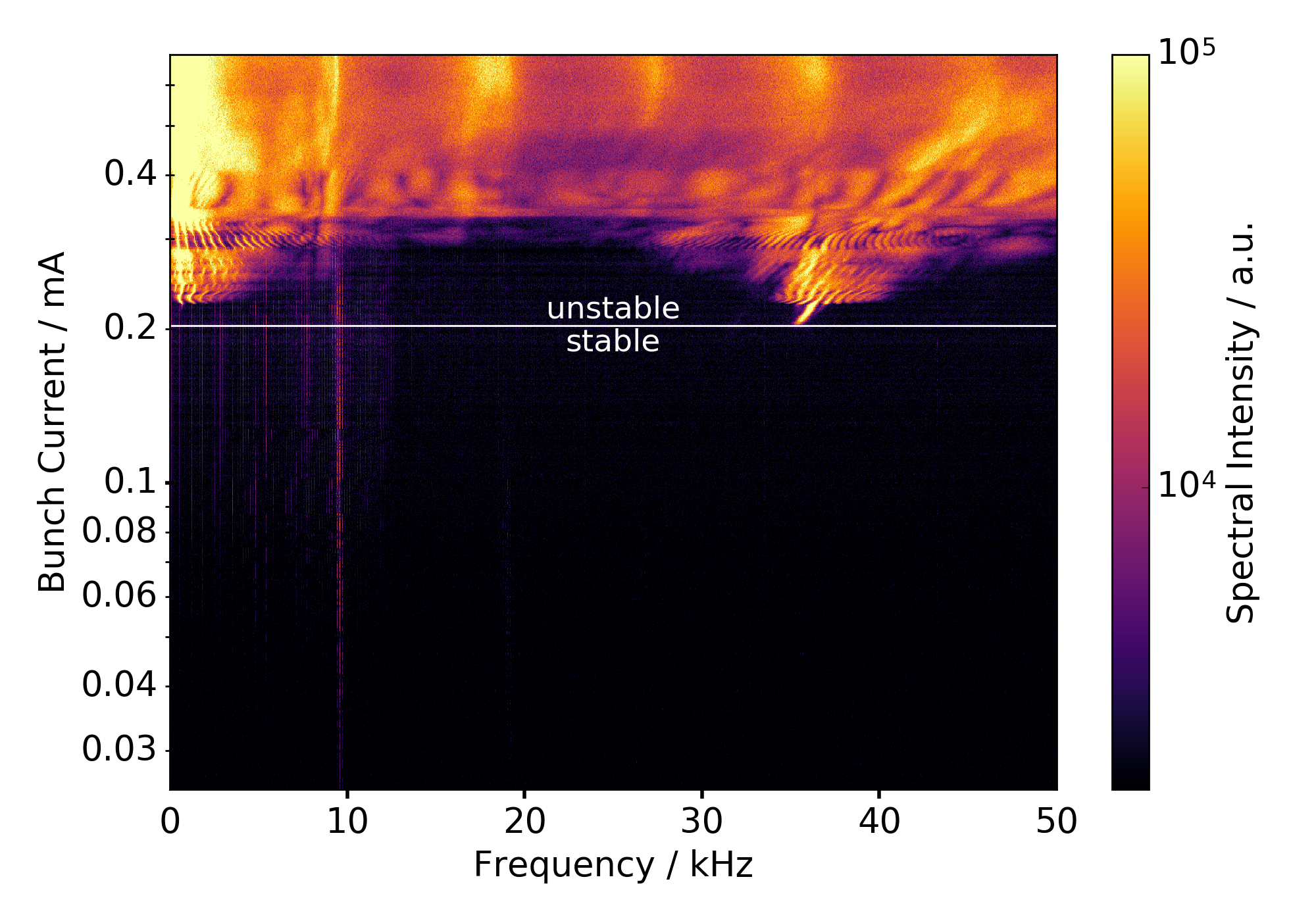}
   \caption{Spectrogram of the fluctuations of the THz intensity as a function of the decaying bunch current, showing the micro-bunching instability. It was obtained in a measurement lasting several hours while the bunch current decreased. No short bunch-length bursting occurs, as the bunch was not compressed strongly enough.}
   \label{fig:decay}
\end{figure}

Figure~\ref{fig:decay} shows the characteristic patterns of the fluctuation frequencies of the emitted THz radiation for different bunch currents. 
During the decay of the bunch current, the instability passes different regimes and ends at the main bursting threshold (in Fig.~\ref{fig:decay} at $\approx$  \unit[0.2]{mA}). 

The combination of a custom filling pattern and a data acquisition system which facilitates the measurement of the THz signal of each bunch individually allows a reduction of the measurement time down to one second \cite{brosi_prab16}. 
This so called snapshot measurement technique was used for measuring the bunch current dependence of the behavior as well as the threshold of the instability.

%The down side of a limited current resolution was attacked by choosing a tight current range around the region of interest. 
%This is visible in the poor current resolution in the upper part of Fig.~\ref{fig:snapshot}.
%
\subsection{Short Bunch-Length Bursting}
\begin{figure}[!tb]
   \includegraphics*[width=0.5\textwidth]{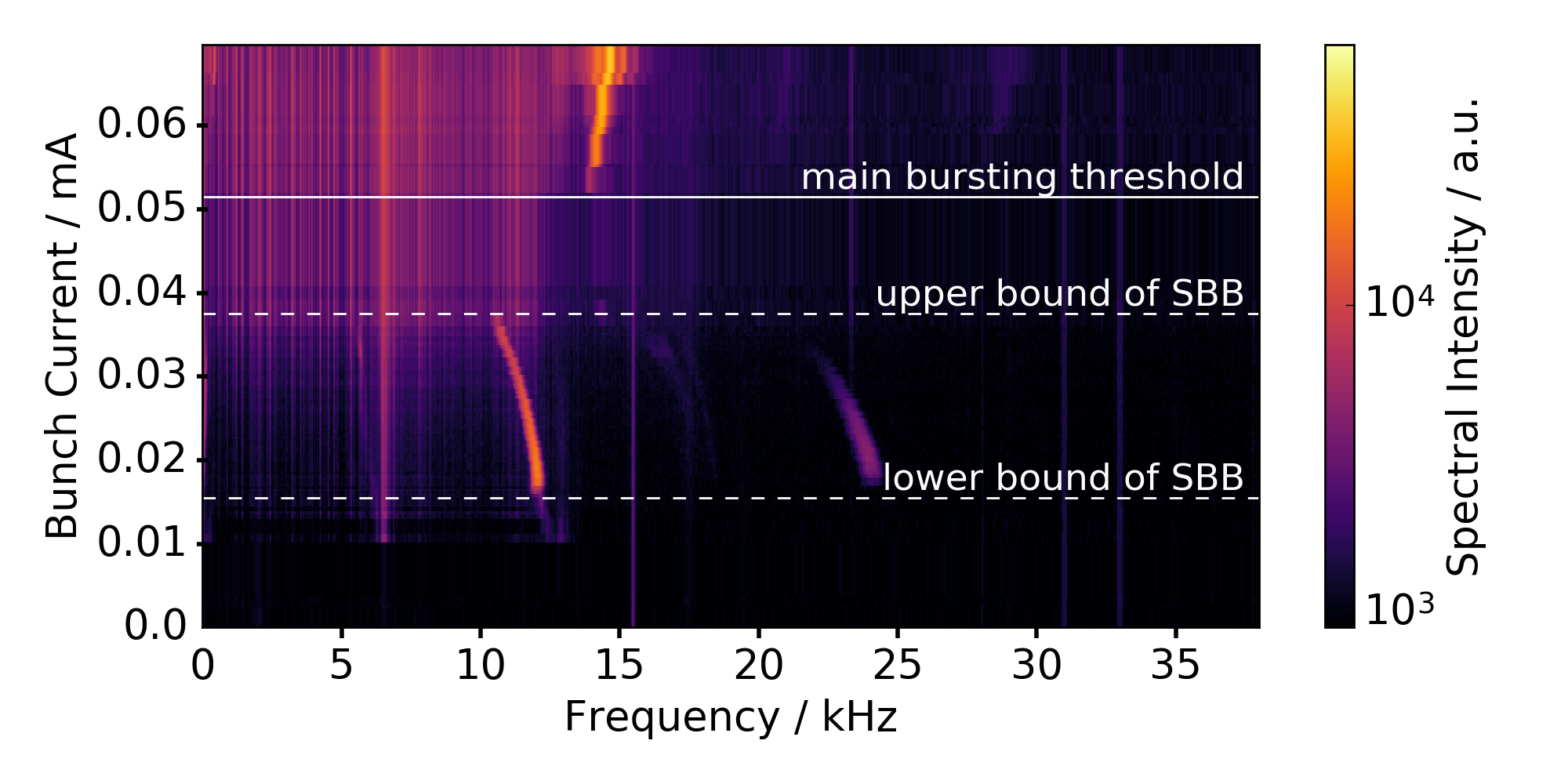}
   \caption{Snapshot spectrogram of the fluctuations of the THz intensity as a function of bunch current for a synchrotron frequency of  \unit[6.55]{kHz}. Below the end of the micro-bunching instability (main bursting threshold) around \unit[0.052]{mA}, a second unstable region is clearly visible between \unit[0.038]{mA} and  \unit[0.016]{mA}. %The limited current resolution in the upper part is due to the snapshot measurement. 
   The current bins were chosen such that a high bunch current resolution in the region of the short bunch-length bursting was achieved. For this measurement approx. $115$ bunches were filled.}
   \label{fig:snapshot}
\end{figure}

%???The short bunch-length bursting occurs below the threshold of the main micro-bunching instability for certain machine settings.??? 
For most machine settings the beam is stable for all currents below the bursting threshold (see Fig.~\ref{fig:decay}). 
Nevertheless, observations show, that at KARA a momentum compaction factor $\alpha_{\mathrm{c}}\le2.64\times10^{-4}$ combined with high RF voltages ($V_{\mathrm{RF}}>\unit[1100]{kV}$) leading to a natural bunch length $\sigma_{\mathrm{z},0}\le\unit[0.723]{mm}\,\hat{=}\, \unit[2.43]{ps}$, an instability occurs again for bunch currents below the main bursting threshold, see Fig.~\ref{fig:snapshot}. 
This spectrogram was obtained by a snapshot measurement within one second. 
To compensate for the limited current resolution of this measurement method, the filling pattern was chosen in such a way that the region of interest with small bunch currents is sampled with a sufficient resolution.
This is visible in the limited bunch current resolution in the upper part of Fig.~\ref{fig:snapshot}.
The spectrogram shows the lower bound of the main bursting and the complete occurrence of the short bunch-length bursting.  
This second region of instability occurred at bunch currents between \unit[0.038]{mA} and \unit[0.016]{mA} for the measured machine settings. % skip last sentence, not interesting?
 
The frequencies of the intensity fluctuations are located below twice the synchrotron frequency ($2\times f_{\mathrm{s}} = 2\times \unit[6.55]{kHz}$ in Fig.~\ref{fig:snapshot}) and approach this frequency with decreasing bunch current. A second frequency line at the first harmonic of the intensity fluctuation is visible (below $4\times f_{\mathrm{s}}$). 

\subsection{Results}

 \begin{figure}[!bt]
   \includegraphics*[width=0.5\textwidth]{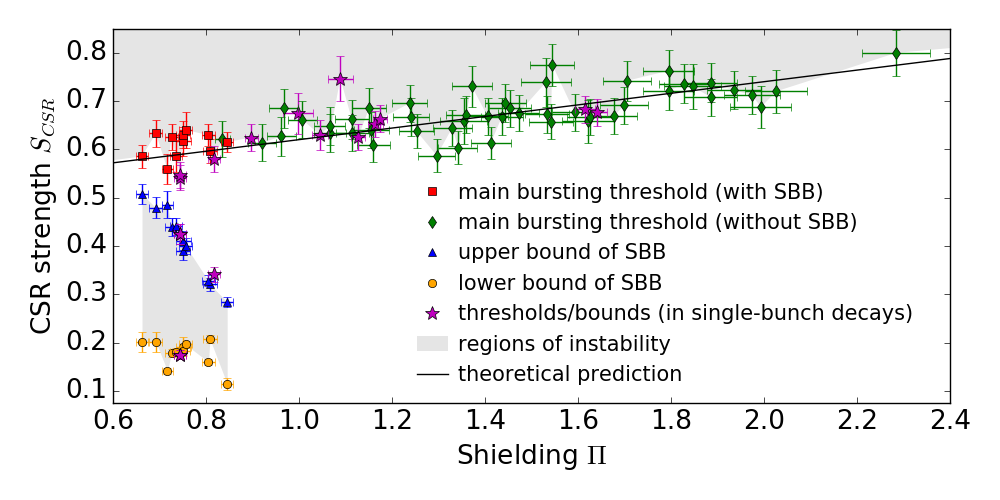}
   \caption{%\textcolor{blue}{
   CSR strength vs. shielding of thresholds from snapshot measurements at different settings of the machine parameters compared to the linear scaling law given by Eq.~\ref{equ:threshold} (line). The lower bound (orange discs) as well as the upper bound (blue triangles) of the short bunch-length bursting (SBB) are shown. The main bursting threshold is shown in red (squares) for machine settings where short bunch-length bursting occurred and in green (diamonds) for settings where it did not occur. The purple stars represent thresholds and bounds which were obtained from a full decay of a single bunch and not from snapshot measurements. The error bars display the one standard deviation uncertainties calculated from the measurement errors.}%} % TODO: update figure, xlabel=Shieling Parameter, ylabel CSR Strength, theoretical prediction -> theoretical calculations
   \label{fig:Scsr_pi}
\end{figure}

Snapshot measurements of the lower current range, similar to Fig.~\ref{fig:snapshot}, were taken for different values of the momentum compaction factor and the natural bunch length by changing the magnet optics as well as the RF voltage. The scanned parameter range reached for $\alpha$ from $9.94\times 10^{-3}$ down to $1.51\times 10^{-3}$ and for $V_\mathrm{RF}$ from $\unit[524]{kV}$ up to $\unit[1500]{kV}$.

The bunch currents at the lower and upper bound of the short bunch-length bursting as well as the main bursting threshold for each measurement are displayed in Fig.~\ref{fig:Scsr_pi} using the dimensionless parameters $S_{\mathrm{CSR}}$ and $\Pi$ (Eqs.~(\ref{equ:pi}) and (\ref{equ:S_csr})) following the notation of \cite{Bane_cai_stupakov2010}. 

Bursting thresholds measured at machine settings, where no short bunch-length bursting occurs (more details see~\cite{brosi_prab16}), show that the main bursting threshold is described by Eq.~(\ref{equ:threshold}) and is independent of the occurrence of short bunch-length bursting. %\todo[inline,color=blue!30]{(Peter) irritating that Eq fits that well, as thresholds were simulated significantly higher than given by eq for $\Pi < 3$. That measurement fits well might be due to additional impedances like resistive wall and/or geometric impedances present in ANKA, which lead to lower thresholds.}

The highest value of the shielding parameter $\Pi$ where the short bunch-length bursting occurs at KARA (right-most red square in Fig.~\ref{fig:Scsr_pi}) is $\Pi_{\mathrm{highest\ SBB}}=0.845\pm0.013$. 
The smallest value of the shielding parameter where the short bunch-length bursting does not occur (left-most green diamond in Fig.~\ref{fig:Scsr_pi}) is at $\Pi_{\mathrm{no\ SBB}}=0.835\pm0.017$, and therefore smaller than $\Pi_{\mathrm{highest\ SBB}}$. 
This small difference is expected and caused by the fact, that the two values ($\Pi_{\mathrm{no\ SBB}}$ and $\Pi_{\mathrm{highest\ SBB}}$) were obtained for different combinations of momentum compaction factor and RF voltage with similar values of $\Pi$, however, different values for $\beta$ ($\beta_{\mathrm{no\ SBB}}=2.59\times 10^{-3}$ and $\beta_{\mathrm{highest\ SBB}}=1.88\times 10^{-3}$). 
As described in \cite{Kuske_ipac17} the range of $\Pi$ where the weak instability occurs is the bigger the smaller $\beta$ is. 
%This is not expected but might be caused by non-linear effects in the magnet optics of the machine, as the two values ($\Pi_{\mathrm{no\ SBB}}$ and $\Pi_{\mathrm{highest\ SBB}}$) were obtained for different combinations of momentum compaction factor and RF voltage with similar values of $\Pi$. % TODO: non-linear stuff raus? ist ja schon im IPAC?

The overall limit agrees within uncertainties with the results obtained by Bane, Stupakov and Cai using their VFP-solver \cite{Bane_cai_stupakov2010}. 
There, the authors observed a dip around $\Pi=0.7$, while the threshold for $\Pi=1$ is again on the theoretical calculated linear scaling law given by Eq.~(\ref{equ:threshold}). 
Values below $\Pi=0.66$ were not accessible for our measurements, which precludes the possibility to check if the short bunch-length bursting vanishes for even smaller values of the shielding parameter, as predicted by calculations in \cite{Bane_cai_stupakov2010}.
 
 \section{Simulations}
 
 \begin{figure}[!bt]
   \includegraphics*[width=0.45\textwidth]{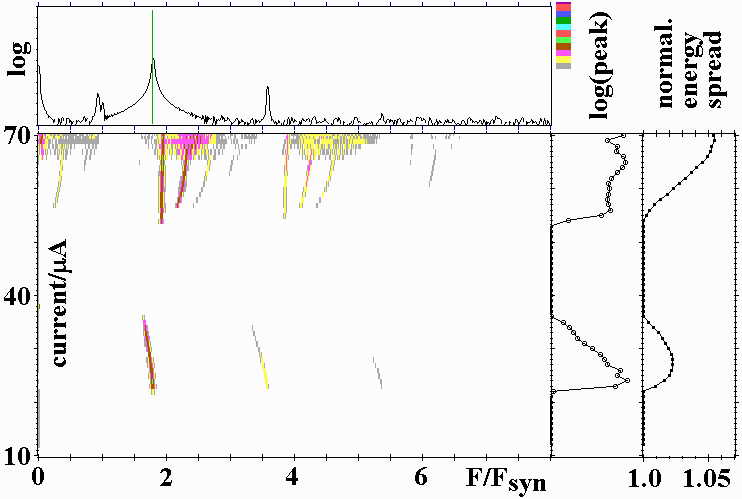}
   \caption{%\textcolor{blue}{
   Simulated spectrogram showing the end of the micro-bunching instability (main bursting threshold) around \unit[54]{$\mu$A} as well as the short bunch-length bursting between \unit[36]{$\mu$A} and \unit[22]{$\mu$A}.
   The frequencies are directly below two and four times the synchrotron frequency, similar to the frequencies observed in the corresponding measurements (see Fig.~\ref{fig:snapshot}).}%}
   \label{fig:sim_spec}
\end{figure}
 
%\textcolor{blue}{
The upper and lower limits in bunch current of the short bunch-length bursting are expected to depend not only on the natural bunch length $\sigma_{\mathrm{z,0}}$ but also on $\beta=1 / \left(2 \pi \, f_{\mathrm{s}}\, \tau_{\mathrm{d}}\right)$ which relates the longitudinal damping time $\tau_{\mathrm{d}}$ and the synchrotron frequency $f_{\mathrm{s}}$. For the measurements presented here, the synchrotron frequency changes due to the different values of the RF-voltage and the momentum compaction factor, while the damping time stays constant. This means that $\beta$ is different for the different measurement points ranging in the presented measurements from %$\beta=1.86\times 10^{-3}$(TODO:decide on correct values)
$\beta=1.13\times 10^{-3}$  to %$\beta=2.86\times 10^{-3}$
$\beta=3.33\times 10^{-3}$. As the simulations in \cite{Bane_cai_stupakov2010} were carried out only for the fixed value $\beta=1.25\times 10^{-3}$, new simulations were done for each measurement point using exactly the parameters of the respective measurement.
The VFP-solver used for these additional simulations is presented in \cite{Kuske_ipac12} and a comparison between the simulation results and measurement done at MLS and BESSY is given in \cite{Kuske_ipac13}.

Figure~\ref{fig:sim_spec} shows a spectrogram calculated from the simulated phase space. 
Similar to the measurements, a second region of instability corresponding to the short bunch-length bursting is visible between \unit[20]{$\mu$A} and \unit[41]{$\mu$A}, well below the main bursting threshold at \unit[54]{$\mu$A}. The dominant frequencies in this instability region are close to two and four times the synchrotron frequency, showing the same structure as the corresponding measurement (Fig.~\ref{fig:snapshot}). 
The simulation also reveals that in the stable area between the short bunch-length bursting and the main bursting threshold, the energy spread equals the natural energy spread and is not increased as it is during the instability (see Fig.~\ref{fig:sim_spec}), confirming the first simulations done in \cite{Bane_cai_stupakov2010}. 

The simulations yield thresholds which are higher by about $10\%$ in comparison to the linear scaling law~\cite{Bane_cai_stupakov2010} as discussed above (see Sec.~\ref{chap:theory}). The overall behavior is in good agreement with the measurements.
The CSR strength at the thresholds obtained from the VFP solver are shown in Fig.~\ref{fig:comp_Scsr_pi} (red triangles) as a function of the shielding parameter.

\section{Comparison}

\begin{figure*}[!tbh]
   \includegraphics*[width=1\textwidth]{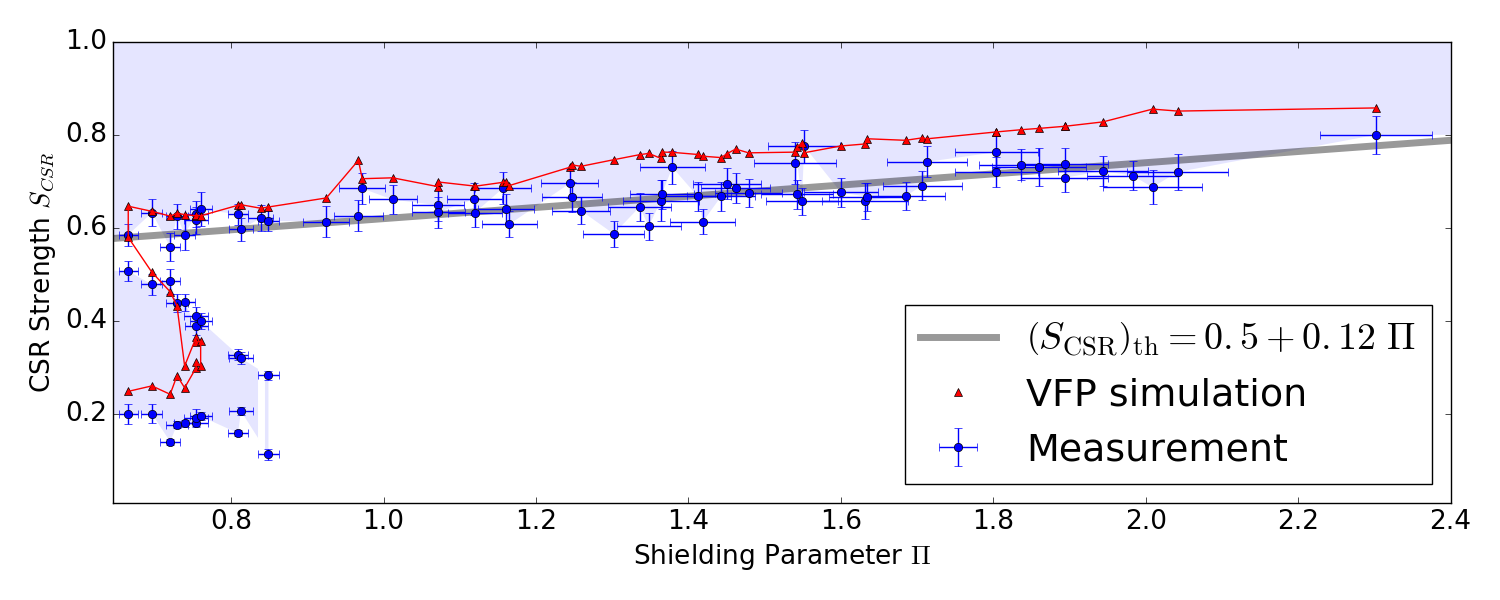}
   \caption{%\textcolor{blue}{
   CSR strength at the bursting thresholds as a function of the shielding parameter for measurements and VFP solver calculations for different machine settings. The measured area of instability is indicated as light blue area and confined by the measured thresholds (blue discs, already shown in Fig.~\ref{fig:Scsr_pi}) with the error bars displaying the standard deviation error of each measurement. The red triangles show the results from the VFP solver calculations at the corresponding machine settings (red line to guide the eye). The gray line indicates the linear scaling law for the main bursting threshold given by Eq.~\ref{equ:S_csr}.%}
   }
   \label{fig:comp_Scsr_pi}
\end{figure*}

%\textcolor{blue}{
Figure~\ref{fig:comp_Scsr_pi} shows again clearly that in the measurement as well as in the new VFP calculations, a range of $\Pi$ exists, where unstable THz emission also occurs below the threshold given by the simple linear scaling law (Eq.~\ref{equ:S_csr}), as already shown by \cite{Bane_cai_stupakov2010}. Our measurements as well as simulations show a stable area between the two regions of instability. 

The range in $\Pi$ as well as in $S_{\mathrm{CSR}}$ where the short bunch-length bursting occurs depends on $\beta$. For the parameters in our measurements, the simulations give an upper limit for the occurrence of short bunch-length bursting of $\Pi=0.76$. %0.76/0.756 TODO: decide on correct value
While the measurements show short bunch-length bursting up to $\Pi_{\mathrm{highest\ SBB}}=0.845\pm0.013$, resulting in a small range of $\Pi$ where short bunch-length bursting is observed by measurements and not in the simulations.  %?(energyspread nicht mal mehr hoch?).\\
Also the lower bound of the short bunch-length bursting in the CSR strength differs slightly between the calculations and the measurements. The measurements show instability at an even lower CSR strength (corresponding to a lower bunch current) than the calculations.
This could be related to the fact that in general the threshold values obtained by the simulation are systematically slightly higher than the ones measured. The average difference is \unit[7]{$\mu$A} for the main threshold current ranging from \unit[40]{$\mu$A} to \unit[400]{$\mu$A}.

Lower values for the thresholds in the measurements can not be explained by too insensitive THz detectors as this would result in an overestimation of the measured thresholds.
Also systematic influences on the measured thresholds due to multi bunch effects in the used snapshot measurements can be excluded, as thresholds measured in pure single-bunch decays agree with the ones from snapshot measurements (see Fig.~\ref{fig:Scsr_pi}).

A small discrepancy consistent with the one observed could be caused due to our measurement method.
For small fluctuations of the machine settings the measurements would give the absolute floor of the corresponding thresholds.
While the simulation gives a value corresponding more to the average threshold for the machine settings, leading to a small discrepancy.
Such fluctuations in the machine could occur in the RF-voltage and the current in the magnets, and thus the magnet optics and the momentum compaction factor.
 
Another potential source for deviations could come from assumptions used in the simulation. For example, the longitudinal damping time, an essential component in the solution of the VFP-equation, was obtained by beam dynamics calculations which did not include CSR. 
Furthermore, as described above, the VFP calculations only consider the simple “parallel plates” model. The small discrepancy between the measured thresholds and the calculated ones could indicate additional impedance contributions.
For example, considering an additional geometric impedance for an aperture like a scraper, leads to a slightly lower simulated threshold \cite{schoenfeldt_diss_18}. Also the impedance of edge radiation is mainly resistive and would lead, if considered in the simulation, to a lower threshold. 
% \todo[inline,color=blue!30]{+Resistive wall? geometric impedances...? Patrik saw small shifts, diss}
 %\todo[inline,color=blue!30]{edge radiation, mainly resistive, lowers threshold....}
Last but not least, a stronger CSR-interaction than expected from the simple circular orbit simulated could be caused by an interaction extending into the straights behind the dipoles.

\section{Summary}
%"good agreement between simple theory and measurements
%• However, systematic differences for the new (single shot)
%measurements
%• Could this be an experimental effect – through the interaction of the
%many bunches?
%• Could this be an indication of a stronger CSR-interaction than
%expected from the simple circular orbit – through an interaction extending into the straights behind the dipoles?"

%\textcolor{gray}{
%The short bunch-length bursting, observed at ANKA for certain machine settings, corresponds to behavior observed in the results of the VFP solver in \cite{Bane_cai_stupakov2010}.  
%This second region of instability occurs below the bursting threshold of the main micro-bunching instability for values of the shielding parameter below $\Pi=0.85$.
%The features and limits agree with the prediction. 
%The difference in the bunch current given by measurement and simulation for the upper bound of the short bunch-length bursting might be explained by a difference in the energy spread (incoherent tune spread?/damping time?) for ANKA and the VFP solver. }

%\textcolor{blue}{
The short bunch-length bursting observed at KARA for certain machine settings, corresponds to the behavior observed in the results of the Vlasov-Fokker-Planck solver calculations first published by Bane, Stupakov and Cai~\cite{Bane_cai_stupakov2010}.  
This second region of instability below the bursting threshold of the main micro-bunching instability occurred in the measurements for values of the shielding parameter smaller than $\Pi=0.85$. As the occurrence of short bunch-length bursting depends not only on the natural bunch length but also on the ratio of the synchrotron frequency and the longitudinal damping time, new VFP solver calculations were performed at the exact conditions of the measurements. This simulation result shows slightly higher values for the thresholds with an average difference of \unit[7]{$\mu$A} but the overall behavior is in good agreement with the measurements. The small deviations might be caused by the floor in the determination of the thresholds from the measurements, which would lead to an underestimation in the presence of small parameter fluctuations of the machine settings.  
Another possible explanation is an additional impedance contribution or a stronger CSR interaction that is not covered by the simple model simulated. The latter could be caused by the interaction of the bunches with their CSR extending into the straights behind the dipoles.
%an interaction that is not covered by the simple model simulated. 
%Another possible explanation is that for small fluctuations in the machine settings the measurements give the lowest threshold value while the VFP calculations would return the threshold for the average settings.%}

\begin{acknowledgments}
We thank Karl Bane for his questions concerning the presence of this second region of instability at KARA.
We would like to thank the infrared group at KARA and in particular M. S\"upfle for their support during the beam times at the Infrared2 beam line. Further, we would like to thank the THz group for inspiring discussions. 
This work has been supported by the German Federal Ministry of Education and Research (Grant No. 05K13VKA), the Helmholtz Association (Contract No. VH-NG-320) and by the Helmholtz International Research School for Teratronics (HIRST).
E. Blomley and J. Gethmann acknowledge the support by the DFG-funded Doctoral School ``Karlsruhe School of Elementary and Astroparticle Physics: Science and Technology''.
M. Brosi, J. Steinmann, and P. Sch\"onfeldt acknowledge the support of the Helmholtz International Research School for Teratronics (HIRST).

\end{acknowledgments}

\appendix*

\bibliography{references_thesis_bibdesk_loaded}

\end{document}